\documentstyle[psfig]{article}

\newcommand{\be}{\begin{equation}}
\newcommand{\ee}{\end{equation}}
\newcommand{\ba}{\begin{eqnarray}}
\newcommand{\ea}{\end{eqnarray}}

\begin{document}

\title{{\bf Harmonically Trapped Quantum Gases}}
\author{M. Grether$^{1}$, M. Fortes$^{2}$, M. de Llano$^{3}$, J.L. del R\'{\i}o$^{4}$%
, \and F.J. Sevilla$^{2}$, M.A. Sol\'{\i}s$^{2}$ and Ariel A. Valladares$%
^{3} $ \\
$^{1}$Facultad de Ciencias, UNAM, Apdo. Postal 70-542,\\
04510 M\'{e}xico, DF, Mexico \\
$^{2}$Instituto de F\'{\i}sica, UNAM, Apdo. Postal 20-364,\\
01000 M\'{e}xico, DF, Mexico \\
$^3$Instituto de Investigaciones en Materiales, UNAM, \\
Apdo. Postal 70-360, 04510 M\'{e}xico, DF, Mexico \\
$^{4}$Universidad Aut\'{o}noma Metropolitana, Unidad\\
Iztapalapa, 09340 M\'{e}xico, DF, Mexico}
\maketitle

\begin{abstract}
We solve the problem of a Bose or Fermi gas in $d$-dimensions trapped by $%
\delta \leq d$ mutually perpendicular harmonic oscillator potentials. From
the grand potential we derive their thermodynamic functions (internal
energy, specific heat, etc.) as well as a generalized density of states. The
Bose gas exhibits Bose-Einstein condensation at a nonzero critical
temperature $T_{c}$ if and only if $d+\delta >2$, and a jump in the specific
heat at $T_{c}$ if and only if $d+\delta >4$. Specific heats for both gas
types precisely coincide as functions of temperature when $d+\delta =2$. The
trapped system behaves like an ideal free quantum gas in $d+\delta $
dimensions. For $\delta =0$ we recover all known thermodynamic properties of
ideal quantum gases in $d$ dimensions, while in 3D for $\delta =$ 1, 2 and 3
one simulates behavior reminiscent of quantum {\it wells, wires }and{\it \
dots}, respectively.


\end{abstract}

{\large 
}

\section{Introduction}

Ultra-cooled bosonic clouds trapped in a harmonic oscillator (HO) external
potential mimic the behavior of bosons confined by realistic potentials as
in opto-magnetic traps in the region of small oscillations. Bose-Einstein
condensation (BEC) has been now observed with $_{37}^{87}$Rb \cite{Ander}, $%
_{11}^{27}$Na \cite{Davis}, $_{3}^{7}$Li \cite{Bradley}, $_{1}^{1}$H \cite
{Fried}, $_{37}^{85}$Rb \cite{Cornish}, $_{2}^{4}$He \cite{Pereira} and $%
_{19}^{41}$K \cite{Mondugno} neutral bosonic atoms, the upper and lower
prefixes being the nuclear mass (number of nucleons in the nucleus) and
proton numbers, respectively.

Trapped quantum gases have been discussed in general by several authors \cite
{Groot}-\cite{Ligare}. The first calculation of the properties of a Bose gas
in an isotropic harmonic trap was reported by de Groot {\it et al}. \cite
{Groot}; Bagnato {\it et al.} \cite{Kleppner} reported theoretical
thermodynamic properties of a Bose gas confined by a generic power-law
potential trap; Ketterle {\it et al. }\cite{Ketterle} and Pathria \cite
{Pathria98} considered the BEC of a finite number of particles confined in a
3D HO and concluded that the thermodynamic-limit approximation is good. For
a review of BEC in trapped dilute Bose gases see Ref. \cite{Dalfovo}.
Trapped Fermi gases have also gained interest as possible precursors of a
paired-fermion condensate at lower temperatures \cite{butts}-\cite{bruun},
and have been studied experimentally in ultracold fermionic clouds, e.g.,
with $_{19}^{40}$K neutral atoms in opto-magnetic traps \cite{MarcoJin}-\cite
{Holland}.

On the other hand, the discovery of the quasi-2D superconductors such as
cuprate \cite{Bednorz}-\cite{poole}{\Huge \ }or the quasi-1D superconductors
such as the organo-metallics (or Bechgaard salts) \cite{jerome}-\cite
{jerome2} have also motivated the study of confinement of quantum gases.

In this paper we describe boson or fermion harmonic-oscillator trapping in
order to better understand these lower-dimensional structures. Since the
system dimensionality modifies the nature of BEC, we seek an exact and
complete solution in the thermodynamic limit to the non-interacting Bose or
Fermi gas problem in $d$-dimensions constrained by a number $\delta $ of
perpendicular HO external potentials. We show that it is possible to map
this problem into that of a free gas but in a higher dimensionality of $%
d+\delta $. For example, confinement \cite{Corcoran}-\cite{corcoran3} in 3D
by a 1D HO potential collapses the system to a quasi-2D ``slab'' reminiscent
of a quantum {\it well.} Confinement by a 2D (or 3D) HO potential leads to a 
{\it wire}-like quasi-1D (or {\it dot}-like quasi-0D) system.

In Sec. 2 we calculate the thermodynamic (or grand) potential for the
non-interacting Bose or Fermi gas in $d$-dimensions with $\delta $
mutually-perpendicular harmonic traps. In Sec. 3 we deduce the thermodynamic
properties of these systems and extract a generalized density of states,\
find their thermodynamic limit, and exhibit their mapping to free gases in
higher dimensions. In Sec. 4 we specialize to a trapped boson gas and obtain
its critical BEC temperature, its condensate fraction and specific heat cusp
or jump. We also summarize our findings for a 3D boson gas trapped by 1, 2
or 3 HO's. In Sec. 5 we specialize to a trapped fermion gas. Sec. 6 contains
our conclusions.

\section{A $d$-dimensional quantum gas trapped by $\protect\delta $ $\leq d$
HOs}

We consider a $d$-dimensional noninteracting boson or fermion gas trapped by 
$\delta =1,2,...,d$ mutually-perpendicular harmonic-oscillator potentials,
the particles otherwise moving freely in the remaining $d-\delta $
directions. The Hamiltonian for a single boson or fermion of mass $m$ is $%
H=\sum_{i=1}^{d}p_{i}^{2}{/2m}+\frac{1}{2}{\,m\,\omega }^{2}{\,}%
\sum_{j=d-\delta +1}^{d}r_{j}^{2}$ and its eigenvalues are 
\begin{equation}
\varepsilon _{\{n_{i},\nu _{j}\}}={\frac{2\pi ^{2}\hbar ^{2}}{mL^{2}}}%
\sum_{i=1}^{d-\delta }n_{i}^{2}+\hbar \omega \sum_{j=1}^{\delta }(\nu
_{j}+1/2)  \label{ener}
\end{equation}
where $L$ is the size of the ``box'' associated with the $d-\delta $ free
dimensions and where $n_{i}=0,\ \pm 1,\;\pm 2,\;...$ while $\nu
_{j}=0,\;1,\;2,\;...$. We may write the grand potential $\Omega (T,V,\mu )$
in generalized form (see p. 134 of \cite{Path}) as \bigskip 
\begin{equation}
\Omega (T,V,\mu )=U-TS-\mu N=\delta _{a,-1}\Omega _{0}-\frac{k_{B}T}{a}%
\sum_{\{{n_{i}},{\nu _{j}\}}}\hbox{ln}[1+ae^{-\beta (\varepsilon
_{\{n_{i},\nu _{j}\}}-\mu )}].  \label{omega}
\end{equation}
Here $V$ $\equiv L^{d-\delta }x_{0}^{2\delta }$ is a confinement volume with 
$x_{0}\equiv \sqrt{{\hbar /m\omega }}$ the oscillator length parameter, $U$
the internal energy, $T$ the absolute temperature, $S$ the entropy, $\mu $
the chemical potential, $N$ the number of particles, $a=-1$ for bosons, $a=1$
for fermions and $a\rightarrow 0$ in the classical case, $\delta $ is the
Kronecker delta function, and $\beta \equiv 1/k_{B}T$ . In the case of a
Bose gas it is convenient to separate out in the sum {(\ref{omega}), }the
lowest energy state from the excited states{. We thus defined }$\Omega
_{0}\equiv -(k_{B}T/a)\hbox{ln}[1+ae^{-\beta (\hbar \omega \delta /2-\mu )}]$
corresponding to the ground state contribution to the grand potential. Using
the logarithm expansion ln$(1+x)=-\sum_{l=1}^{\infty }(-x)^{l}/l${\ valid
for }$\left| x\right| {<1}$, {(\ref{omega}) becomes} 
\begin{eqnarray}
\Omega (T,V,\mu ) &=&\delta _{a,-1}\Omega _{0}+\frac{k_{B}T}{a}%
\sum_{\{n_{i},\nu _{j}\}}\sum_{{\it l}=1}^{\infty }{\frac{(-{\it a}e^{-\beta
(\varepsilon _{\{n_{i},\nu _{j}\}}-\mu )})^{{\it l}}}{{\it l}}}  \nonumber \\
&=&\delta _{a,-1}\Omega _{0}-\frac{k_{B}T}{a}\sum_{{\it l}=1}^{\infty }{%
\frac{(-{\it a}e^{\beta \mu })^{{\it l}}}{{\it l}}}\sum_{\{n_{i},\nu
_{j}\}}e^{-\beta {\it l}\left[ (\hbar ^{2}2\pi ^{2}/mL^{2}{)}%
\sum_{i=1}^{d-\delta }n_{i}^{2}+\hbar \omega \sum_{j=1}^{\delta }(\nu
_{j}+1/2)\right] }.
\end{eqnarray}
Next, consider only the excited states as the ground state will be treated
separately for the boson gas. In the continuous limit where $\hbar
^{2}/mL^{2}\ll k_{B}T$ and $\hbar \omega \ll k_{B}T$, the summations over $%
n_{i}$ and $\nu _{j}$ can be approximated by integrals, namely ${\ }\sum_{%
{\bf n}}\ \ {\longrightarrow \ \ }(2{\tt s}+1)\int d^{d}n_{i}$. Thus 
\begin{eqnarray}
\Omega (T,V,\mu ) &=&\delta _{a,-1}\Omega _{0}-{\frac{(2{\tt s}+1)k_{B}T}{%
{\it a}}}\sum_{l=1}^{\infty }{\frac{(-{\it a}e^{\beta \mu })^{{\it l}}}{{\it %
l}}}\int_{-\infty }^{\infty }dn_{1}\ e^{{-\beta }{\it l}(\hbar ^{2}2\pi
^{2}/mL^{2}{)}n_{1}^{2}}  \nonumber \\
&&\times \int_{-\infty }^{\infty }dn_{2}\ e^{{-\beta }{\it l}(\hbar ^{2}2\pi
^{2}/mL^{2}{)}n_{2}^{2}}...\int_{-\infty }^{\infty }dn_{d-\delta }\ e^{{%
-\beta }{\it l}(\hbar ^{2}2\pi ^{2}/mL^{2}{)}n_{d-\delta }^{2}}  \nonumber \\
&&\times \int_{0}^{\infty }d\nu _{1}\ e^{{-\beta }{\it l}\hbar \omega (\nu
_{1}+1/2)}...\int_{0}^{\infty }d\nu _{\delta }\ e^{{-\beta }{\it l}\hbar
\omega (\nu _{\delta }+1/2)}.
\end{eqnarray}
where ${\tt s}$ is the particle spin, with fermions having ${\tt s}${\tt \ }$%
{\tt =}1/2$ and bosons ${\tt s}${\tt \ }$=0$. The integrals are elementary
and give 
\begin{equation}
\Omega (T,V,\mu )=\delta _{a,-1}\Omega _{0}-{\frac{2{\tt s}+1}{{\it a}}\beta
^{-[\left( d+\delta \right) /2+1]}}\left( {\frac{mL^{2}}{2\pi \hbar ^{2}}}%
\right) ^{(d-\delta )/2}\left( \hbar \omega \right) ^{-\delta
}\sum_{l=1}^{\infty }{\frac{[-{\it a}e^{\beta (\mu -\delta \hbar \omega
/2)}]^{{\it l}}}{{\it l}^{\left( d+\delta \right) /2+1}}.}  \label{omega4}
\end{equation}
The infinite sum is expressible in terms of the{\it \ polylogarithm function}
$Li_{\sigma }(z)$ \cite{Math}, since 
\begin{equation}
-{\it a}Li_{\sigma }(-{\normalsize a}z)\equiv {\frac{1}{\Gamma (\sigma )}}%
\int_{0}^{\infty }dx{\frac{x^{\sigma -1}}{z^{-1}e^{x}+{\it a}}}=-\frac{{1}}{a%
}\sum_{l=1}^{\infty }{\frac{(-{\it a}z)^{{\it l}}}{{\it l}^{\sigma }}.}
\label{poly}
\end{equation}
For $\sigma \geq 1$ this reduces to Bose-Einstein{\it \ }(BE) integrals $%
g_{\sigma }(z)$ when $a=-1$ and to Fermi-Dirac (FD) integrals $f_{\sigma
}(z) $ when $a=1$, as defined in Appendix D and E of Ref. \cite{Path}, and $%
z\equiv e^{\beta \mu }$ is the fugacity. Using {(\ref{poly}) in (\ref{omega4}%
) }leaves \bigskip 
\begin{equation}
\Omega \left( T,V,\mu \right) =\delta _{a,-1}\Omega _{0}-\frac{{1}}{a}{\frac{%
A_{d+\delta }}{\beta ^{\left( d+\delta \right) {/2}+1}}Li_{\left( d+\delta
\right) {/2}+1}(-{\it a}z_{1})}  \label{TGP}
\end{equation}
where 
\begin{equation}
A_{d+\delta }\equiv {\frac{2{\tt s}+1}{(\hbar \omega )^{\delta }}}\left(
mL^{2}/2\pi \hbar ^{2}\right) ^{(d-\delta )/2}  \label{Ad}
\end{equation}
and 
\begin{equation}
z_{1}\equiv ze^{-\beta \delta \hbar \omega /2}=e^{\beta (\mu -\delta \hbar
\omega /2)}.  \label{fug}
\end{equation}

\section{Thermodynamic properties}

From (\ref{TGP}) it is possible to find the thermodynamic properties for a
monatomic gas using the relation 
\begin{equation}
d\Omega =-SdT-PdV-Nd\mu .  \label{relfun}
\end{equation}
In this representation the grand potential $\Omega (T,V,\mu )$\ $=-PV$ is
the fundamental relation leading to all the thermodynamic properties of the
system since 
\begin{equation}
N=-\left( {\frac{\partial \Omega }{\partial \mu }}\right) _{T,V},{\ \ \ \ \
\ \ }S=-\left( {\frac{\partial \Omega }{\partial T}}\right) _{V,\mu },\quad
\ \ \ \ P=-\left( {\frac{\partial \Omega }{\partial V}}\right) _{T,\mu }=-%
\frac{\Omega }{V}.  \label{edos}
\end{equation}
Using (\ref{TGP}) and (\ref{edos}) the particle number is given by 
\begin{equation}
N=-{\frac{A_{d+\delta }}{{\it a}\beta ^{\left( d+\delta \right) {/2}}}%
Li_{\left( d+\delta \right) {/2}}(-{\it a}z_{1})},\quad \quad   \label{num1}
\end{equation}
where we used the relation 
\begin{equation}
\left( \frac{\partial {Li_{\left( d+\delta \right) {/2+1}}(-{\it a}z_{1})}}{%
\partial \mu }\right) _{T,V}=\beta \ {Li_{\left( d+\delta \right) {/2}}(-%
{\it a}z_{1})}.
\end{equation}
The entropy follows on substituting (\ref{TGP}) in\ the first equation of (%
\ref{edos}),\ giving 
\begin{equation}
S/k_{B}=-[({d+\delta )/}2+1]{\frac{A_{d+\delta }}{{\it a}\beta ^{\left(
d+\delta \right) {/2}}}Li_{\left( d+\delta \right) {/2}+1}(-{\it a}z_{1})}-N{%
\ }{\rm ln}{\ }z_{1},  \label{s0}
\end{equation}
where\ we used the number equation (\ref{num1}) and the relation 
\begin{equation}
\left( \frac{\partial {Li_{\left( d+\delta \right) {/2+1}}(-{\normalsize a}%
z_{1})}}{\partial T}\right) _{V,\mu }=\frac{1}{z_{1}}\left( {\frac{\partial
z_{1}}{\partial T}}\right) _{V,\mu }{Li_{\left( d+\delta \right) {/2}}(-{\it %
a}z_{1})}.
\end{equation}
Thus (\ref{s0}) becomes
\begin{equation}
S/Nk_{B}=\frac{[({d+\delta )/}2+1]{Li_{\left( d+\delta \right) {/2}+1}(-{\it %
a}z_{1})}}{{Li_{\left( d+\delta \right) {/2}}(-{\it a}z_{1})}}-{\rm ln}{\ }%
z_{1}.  \label{S}
\end{equation}
The internal energy is obtained from (see p. 159 of Ref. \cite{Path}) 
\begin{equation}
U(T,V)=-k_{B}T^{2}\left[ {\frac{\partial }{\partial T}}\left( {\frac{\Omega 
}{k_{B}T}}\right) \right] _{V,\ z}.  \label{U-1}
\end{equation}
Substituting (\ref{TGP}) here we find that
\begin{equation}
U(T,V)=N{\frac{\hbar \omega \delta }{2}}-\frac{d+\delta }{2}{\Omega ,}
\label{u3}
\end{equation}
and since $\Omega =-PV$ then 
\begin{equation}
{PV}{\ \ }{=}{\ }\frac{2}{d+\delta }(U-N\hbar \omega \delta /2{)}.
\end{equation}
\bigskip Using (\ref{Ad}) and (\ref{num1}) the internal energy (\ref{u3})
can be rewritten as
\begin{equation}
\frac{U(T,V)}{Nk_{B}T}=\left[ \beta {\frac{\hbar \omega \delta }{2}}+{\frac{%
d+\delta }{2}}\frac{{Li_{\left( d+\delta \right) {/2}+1}(-{\it a}z_{1})}}{{%
Li_{\left( d+\delta \right) {/2}}(-{\it a}z_{1})}}\right] .  \label{u-3}
\end{equation}
The specific heat at constant volume $C_{V}$ then follows from 
\begin{equation}
C_{V}=\left[ {\frac{\partial }{\partial T}}U(T,V)\right] _{N,V}  \label{dcv}
\end{equation}
and gives 
\begin{equation}
\frac{C_{\mbox v}}{Nk_{B}}=\frac{d+\delta }{2}\left[ (\frac{d+\delta }{2}+1)%
\frac{{Li_{\left( d+\delta \right) {/2}+1}(-{\it a}z_{1})}}{{Li_{\left(
d+\delta \right) {/2}}(-{\it a}z_{1})}}-\frac{d+\delta }{2}\frac{{Li_{\left(
d+\delta \right) {/2}}(-{\it a}z_{1})}}{{Li_{\left( d+\delta \right) {/2}%
-1}(-{\it a}z_{1})}}\right]  \label{CV}
\end{equation}
where we have used the relation 
\begin{equation}
\frac{1}{z_{1}}\left( {\frac{\partial z_{1}}{\partial T}}\right)
_{N,V}=-k_{B}\beta \frac{d+\delta }{2}\frac{{Li_{\left( d+\delta \right) {/2}%
}(-{\it a}z_{1})}}{{Li_{\left( d+\delta \right) {/2}-1}(-{\it a}z_{1})}}
\end{equation}
which can be extracted from the (vanishing) derivative with respect to $T$
of the number equation (\ref{num1}). Since $z_{1}\equiv e^{\beta (\mu
-\delta \hbar \omega /2)}%
\smash {\
\mathop{\relbar\joinrel\longrightarrow}\limits_{T\to \infty}\ \ }0$, (\ref
{poly}) then implies that $-{\it a}{Li_{\sigma }(-{\it a}z_{1})}\rightarrow
z_{1}$, with $\sigma =(d+\delta )/2-1$, $(d+\delta )/2$ or $(d+\delta )/2+1$%
. Then ($\ref{CV}$) reduces to
\begin{equation}
\frac{C_{\mbox v}}{Nk_{B}}\smash {\
\mathop{\relbar\joinrel\longrightarrow}\limits_{T\to \infty}\ \ }\frac{%
d+\delta }{2}\left[ 1+a\frac{z_{1}}{2^{\left( d+\delta \right) /2+1}}\left(
1-\frac{d+\delta }{2}\right) \right] ,  \label{cvinf}
\end{equation}
which for $\delta =3$\ gives the classical Dulong-Petit law\ for $%
T\rightarrow \infty $ or $z_{1}=0,$ while for $\delta =0$ we obtain the
classical limit for ideal gases of \ bosons or fermions. The first
correction to unity in ($\ref{cvinf}$) for $d+\delta <2$ is clearly negative
for $a=-1$ (bosons) and positive for $a=+1$ (fermions), while for $d+\delta
>2$ it is precisely the opposite. Thus we obtain known results obtained for
ideal gases for $\delta =0$ [Refs. \cite{Ziff} (bosons), \cite{Ariel}
(fermions)].

We now recover the results obtained in Refs. \cite{may}-\cite{pathriaA}
dealing with the equivalence of the specific heat as a function of $T$ of
ideal Bose and Fermi gases in two dimensions. Here this equivalence is
obtained more generally for $d+\delta =2$. If both gases are at the same
temperature and have the same number density $n_{B}=n_{F},$ where$\
n_{B}\equiv N_{B}/V$ is the Bose and $n_{F}$ $\equiv N_{F}/V$ is the Fermi
density, taking $d+\delta =2$ in (\ref{num1}) gives 
\begin{equation}
{\ }n_{B}={\frac{A_{2}}{V}}\frac{{Li_{1}(z_{1B})}}{\beta }=-{\frac{A_{2}}{V}}%
\frac{{Li_{1}(-z_{1F})}}{\beta }=n_{F},  \label{denbf}
\end{equation}
where\ as before $V$ was defined just below (\ref{omega}), $z_{1B}\equiv
e^{\beta (\mu _{B}-\hbar \omega \delta /2)}$\ and $z_{1F}\equiv e^{\beta
(\mu _{F}-\hbar \omega \delta /2)}$ are the fugacities with $\mu _{B}$ and $%
\mu _{F}$ the chemical potentials for bosons and fermions, respectively.
Using Landen's relations \cite{lee} the polylogarithm functions ${Li_{\sigma
}({\it z})}$ satisfy ${Li_{1}({\it x})=-Li_{1}({\it y})}\ $and$\ {Li_{2}(%
{\it x})=-Li_{2}({\it y})-1/2}\left[ {Li_{1}({\it y})}\right] ^{2},$ where $%
x\rightarrow y$ satisfy the Euler transformation $y\equiv -x/(1-x)$ with $x$
real $<1$. Substituting these relations in (\ref{denbf}), we obtain 
\begin{equation}
z_{1F}=z_{1B}/\left( 1-z_{1B}\right) .  \label{fug BF}
\end{equation}
The energy of the Bose gas $U(T,V)_{B}$ taking $a=-1$ in (\ref{u-3}) with $%
d+\delta =2$, is 
\begin{equation}
\frac{U(T,V)_{B}}{Nk_{B}T}=\left[ \beta {\frac{\hbar \omega \delta }{2}}+%
\frac{{Li_{2}(z_{1B})}}{{Li_{1}(z_{1B})}}\right] .  \label{u2B}
\end{equation}
Substituting (\ref{fug BF}) in (\ref{u2B}) we obtain 
\begin{eqnarray}
\frac{U(T,V)_{B}}{N} &=&\left[ {\frac{\hbar \omega \delta }{2}}+\beta ^{-1}%
\frac{{Li_{2}(-z_{1F})}}{{Li_{1}(-z_{1F})}}+1/2\beta ^{-1}{Li_{1}(-z_{1F})}%
\right]  \nonumber \\
&=&\left[ \frac{U(T,V)_{F}}{N}+1/2\beta ^{-1}{Li_{1}(-z_{1F})}\right] ,
\label{ubf}
\end{eqnarray}
where $U(T,V)_{F}$ is the Fermi gas energy. Substituting (\ref{denbf}) in (%
\ref{ubf}) the last term in (\ref{ubf}) is proportional to $n_{F}$. Hence,
the energies of the Bose and Fermi gases differ only by a $T$-independent
term and so, from (\ref{dcv}), the specific heats for boson and fermion
gases precisely coincide when $d+\delta =2,$ or 
\begin{equation}
\left[ C_{\mbox v}(N,T)\right] _{B}=\left[ C_{\mbox v}(N,T)\right] _{F}.
\end{equation}

\subsection{Mapping to higher-$d$ and equivalent mass}

\bigskip Using (\ref{poly}) and (\ref{fug}), equation (\ref{num1}) can be
rewritten as 
\begin{eqnarray}
N &=&{\frac{A_{d+\delta }}{\Gamma \left[ (d+\delta )/2\right] }}%
\int_{0}^{\infty }d\varepsilon {\frac{\varepsilon ^{(d+\delta )/2-1}}{%
z_{1}^{-1}e^{\beta \varepsilon }+{\it a}}=\frac{A_{d+\delta }}{\Gamma \left( %
\left[ d+\delta \right] /2\right) }}\int_{\hbar \omega \delta /2}^{\infty
}d\varepsilon {\frac{(\varepsilon -\hbar \omega \delta /2)^{(d+\delta )/2-1}%
}{e^{\beta (\varepsilon -\mu )}+{\it a}}}  \nonumber \\
{} &\equiv &{}\int_{\hbar \omega \delta /2}^{\infty }d\varepsilon {\cal N}%
\left( \varepsilon \right) n\left( \varepsilon \right) ,  \label{Num}
\end{eqnarray}
where \ $n\left( \varepsilon \right) =\left[ e^{\beta \left( \varepsilon
-\mu \right) }+a\right] ^{-1}$\ is the BE ($a=-1$) or FD ($a=+1$)
distribution, and ${\cal N}\left( \varepsilon \right) $ is the density of
states (DOS). Substituting $A_{d+\delta }$ from (\ref{Ad}) into (\ref{Num})
we identify this {\it generalized} DOS ${\cal N}(\varepsilon )$ as 
\begin{equation}
{\cal N}(\varepsilon )=\left( 2{\tt s}+1\right) {\left( {\frac{2\pi \hbar }{%
m\omega L^{2}}}\right) ^{\delta }}\left( \frac{{mL}^{2}}{2\pi \hbar ^{2}}%
\right) ^{\left( d+\delta \right) /2}{\frac{(\varepsilon -\hbar \omega
\delta /2)^{(d+\delta )/2-1}}{\Gamma \left( \left[ d+\delta \right]
/2\right) }}.  \label{DOS}
\end{equation}
If$\ \delta =0$\ we recover the DOS for a free gas confined in a ``box'' of
sides $L$ 
\begin{equation}
{\cal N}_{0}(\varepsilon )=\left( 2{\tt s}+1\right) \left( \frac{{mL}^{2}}{%
2\pi \hbar ^{2}}\right) ^{d/2}{\frac{\varepsilon ^{d/2-1}}{\Gamma \left(
d/2\right) }}.  \label{dosfree}
\end{equation}
Comparing (\ref{DOS}) with (\ref{dosfree}) in $\left( d+\delta \right) $%
-dimensions 
\begin{equation}
{\cal N}_{0}(\varepsilon )=\left( 2{\tt s}+1\right) \left( \frac{{mL}^{2}}{%
2\pi \hbar ^{2}}\right) ^{\left( d+\delta \right) /2}{\frac{\varepsilon
^{\left( d+\delta \right) /2-1}}{\Gamma \left( \left[ d+\delta \right]
/2\right) },}  \label{d0}
\end{equation}
we observe that except for the (negligible) zero-point energy of $\hbar
\omega \delta /2$, (\ref{DOS}) and (\ref{d0}) are identical if in (\ref{DOS}%
) an{\it \ equivalent }particle mass $m^{\ast }$ defined by
\begin{equation}
m^{\ast }=\left( h/\omega L^{2}\right) ^{2\delta /(d+\delta )}m^{(d-\delta
)/(d+\delta )}  \label{mequiv}
\end{equation}
is introduced. Then 
\begin{equation}
{\cal N}(\varepsilon )=\left( 2{\tt s}+1\right) {(}m^{\ast }L^{2}/2\pi \hbar
^{2})^{\left( d+\delta \right) /2}\frac{\varepsilon ^{\left( d+\delta
\right) /2-1}}{\Gamma \left( \left[ d+\delta \right] /2\right) }.
\end{equation}
In general, therefore, the effect of trapping a quantum gas renormalizes the
particle mass $m\rightarrow m^{\ast }$ in accordance with (\ref{mequiv}) and
increases the dimensionality $d\rightarrow d+\delta $ by the number of
oscillators.

\subsection{Thermodynamic limit}

\bigskip\ Substituting the coefficient $A_{d+\delta }$\ from (\ref{Ad}) into
(\ref{Num}) gives 
\begin{equation}
N=\left( {2{\tt s}+1}\right) \left( \frac{{m}}{2\pi \hbar ^{2}}\right)
^{\left( d+\delta \right) /2}{\left( {\frac{2\pi \hbar }{m\omega }}\right)
^{\delta }\frac{x_{0}^{-2\delta }V}{\Gamma \left( \left[ d+\delta \right]
/2\right) }}\int_{\hbar \omega \delta /2}^{\infty }d\varepsilon {\frac{%
(\varepsilon -\hbar \omega \delta /2)^{(d+\delta )/2-1}}{e^{\beta
(\varepsilon -\mu )}+{\it a}}}  \label{Num1}
\end{equation}
the volume $V$ being defined just below (\ref{omega}). The proper
thermodynamic limit then holds if $N\rightarrow \infty ,$ $L\rightarrow
\infty ,$ $\omega \rightarrow 0$\ while keeping the ratio $N/V=N/L^{d-\delta
}x_{0}^{2\delta} \propto N\omega ^{\delta }/L^{d-\delta }$ = constant.
This result was obtained for $d=3$ and $\delta =3$ in Ref. \cite{Dalfovo}.
For a free gas, i.e. $\delta =0$, we recover the usual thermodynamic limit,
i.e., $N\rightarrow \infty ,L\rightarrow \infty $, while $N/L^{d}=$ constant$%
.$

\section{Trapped bosons}

In this section we study a system of $N$ noninteracting bosons in $d$
dimensions trapped by $\delta $ $(\leq d)$ mutually-perpendicular harmonic
oscillators, and otherwise free in the remaining $d-\delta $ directions. Let
the boson number be 
\begin{equation}
N=N_{0}(T)+N_{{\bf k}>0}(T)  \label{N-1}
\end{equation}
were $N_{0}(T)=-\left( \partial \Omega _{0}/\partial \mu \right) _{T,V}$ \
is the number of bosons in the lowest energy state, with $\Omega _{0}$
defined just below (\ref{omega}), while $N_{{\bf k}>0}(T)$ is given by (\ref
{num1}) with $a=-1.$ Thus 
\begin{equation}
N=N_{0}(T)+\frac{A_{d+\delta }}{\beta ^{(d+\delta )/2}}g_{(d+\delta
)/2}(z_{1}),  \label{Ne1}
\end{equation}
where from (\ref{poly})$\ $we introduce the Bose function $g_{\sigma }(z)$
which for $z=1$ and $\sigma >1$ is identical to the Riemann Zeta function $%
\zeta (\sigma )$.

Since for $T>T_{c}$, $N_{0}(T)$ is negligible compared with $N$, while for $%
T<T_{c}$, $N_{0}(T)$ is a sizeable fraction of $N,$ at $T=T_{c},$ $%
N_{0}(T_{c})\simeq 0$. The critical temperature $T_{c}$ of BEC is found from
the condition $N_{{\bf k}>0}(T_{c},z_{1}=1)\simeq N$, \ so that (\ref{Ne1})
leads to 
\begin{equation}
k_{B}T_{c}=\left[ \frac{N}{A_{d+\delta }g_{(d+\delta )/2}(1)}\right]
^{2/(d+\delta )}.  \label{Tc}
\end{equation}
From (\ref{poly}) the infinite series $g_{\sigma }(1)$ diverges for $\sigma
\leq 1$ implying from (\ref{Tc}) that BEC will occur with critical
temperature $T_{c}\neq 0$ if and only if $(d+\delta )/2>1.$ For $\delta =0$
and $d=3$ with{\Huge \ }$n\equiv N/L^{3}$ (\ref{Tc}) reduces to the familiar
formula $T_{c}\simeq 3.31\hbar ^{2}n^{2/3}/mk_{B}$ of ``ordinary'' BEC,
since $g_{3/2}(1)=\zeta (3/2)\simeq 2.612$. On the other hand, substituting $%
\delta =3$ and $d=3$ in (\ref{Tc}) and (\ref{Ad}), we recover the result
obtained in \cite{Dalfovo}, 
\begin{equation}
k_{B}T_{c}\simeq 0.94\hbar \omega N^{1/3}.  \label{tc3d}
\end{equation}
Ensher {\it et al., }\cite{Ensher} compared the experimental $T_{c}$
obtained for real traps with the theoretical value (\ref{tc3d}) and found
good agreement.

From (\ref{Ne1}) and (\ref{Tc}) we obtain the condensate fraction, 
\begin{equation}
N_{0}(T)/N=1-N_{{\bf k}>0}(T)/N\left( T_{c}\right) =1-\left( T/T_{c}\right)
^{(d+\delta )/2}.
\end{equation}
The specific heat follows from (\ref{CV}) for $a=-1$ and from $-{a}%
Li_{\sigma }(-az)\left[ \sigma ,z_{1}\right] =g_{\sigma }(z_{1}).$ We obtain
for $T>T_{c}$ 
\begin{equation}
\frac{C_{\mbox v}}{Nk_{B}}=\frac{d+\delta }{2}\left[ (\frac{d+\delta }{2}+1)%
\frac{g_{(d+\delta )/2+1}(z_{1})}{g_{(d+\delta )/2}(z_{1})}-\frac{d+\delta }{%
2}\frac{g_{(d+\delta )/2}(z_{1})}{g_{(d+\delta )/2-1}(z_{1})}\right] ,
\label{cvbt}
\end{equation}
while for $T\leq T_{c}$, $z_{1}=1$ it follows directly from (\ref{u-3}) and (%
\ref{dcv}) that 
\begin{equation}
\frac{C_{\mbox v}}{Nk_{B}}=\frac{d+\delta }{2}(\frac{d+\delta }{2}+1)\left(
T/T_{c}\right) ^{(d+\delta )/2}\frac{g_{(d+\delta )/2+1}(1)}{g_{(d+\delta
)/2}(1)}.  \label{CVb}
\end{equation}
The specific heat jump at $T_{c}$ is then 
\begin{equation}
\frac{\Delta C_{\mbox v}}{Nk_{B}}=\frac{C_{V}(T_{c}^{-})-C_{V}(T_{c}^{+})}{%
Nk_{B}}=\left( \frac{d+\delta }{2}\right) ^{2}\frac{g_{(d+\delta )/2}(1)}{%
g_{(d+\delta )/2-1}(1)},  \label{jump}
\end{equation}
and, since $g_{\sigma }(1)$ diverges for $\sigma \leq 1$, will be nonzero if
and only if $(d+\delta )/2>2.$

The entropy for $a=-1$ follows from (\ref{s0}). Since $-{a}Li_{\sigma }(-az)%
\left[ \sigma ,z_{1}\right] =g_{\sigma }(z_{1})$ and using (\ref{num1}), in
terms of the critical temperature $T_{c}$ it becomes
\begin{equation}
S/Nk_{B}=[({d+\delta )/}2+1]\left( T/T_{c}\right) ^{(d+\delta )/2}\frac{%
g_{(d+\delta )/2+1}(z_{1})}{g_{(d+\delta )/2}(1)}-{\normalsize \ }{\rm ln}{\ 
}z_{1}.  \label{sb}
\end{equation}
For $T\leq T_{c},$ $z_{1}=1,$ so that this becomes

\[
S/Nk_{B}=[({d+\delta )/}2+1]\left( T/T_{c}\right) ^{(d+\delta )/2}\frac{%
g_{(d+\delta )/2+1}(1)}{g_{(d+\delta )/2}(1)}%
\smash {\
\mathop{\relbar\joinrel\longrightarrow}\limits_{T\to 0}\ \ }0 
\]
which complies with the third law of thermodynamics.

For 3D bosons trapped by 1, 2 or 3 harmonic oscillators we summarize our
results in Table 1. Since for $z_{1}=1$ the series$\ g_{\sigma }(z_{1})$ for 
$\sigma >1$ coincides with $\zeta (\sigma ),$ we require the following
values: $\zeta (3/2)\simeq 2.612,$ $\zeta (2)\simeq 1.645,$ $\zeta
(5/2)\simeq 1.341,$ $\zeta (3)\simeq 1.202,$ $\zeta (3/2)\simeq 1.127,$ and $%
\zeta (4)\simeq 1.082$. In Fig. 1 we show the condensate fraction for $%
\delta =1,2\ $and$\ 3.$ In Fig. 2 are shown their internal energy; specific
heat at constant volume (having a jump discontinuity if and only if $%
d+\delta >4$); entropy and chemical potential. \bigskip

\begin{center}
\begin{tabular}{|c||c|c|c|}
\hline
$\ \delta $ & $3$ & $2$ & $1$ \\ \hline\hline
${\cal N}(\varepsilon )$ & ${{\frac{1}{2}}(\hbar \omega )^{-3}(\varepsilon -{%
\frac{3}{2}}\hbar \omega )^{2}}$ & ${{\frac{2^{3/2}}{3}}{\frac{L}{\pi x_{0}}}%
(\hbar \omega )^{-5/2}(\varepsilon -\hbar \omega )^{3/2}}$ & ${\frac{L^{2}}{%
2\pi x_{0}^{2}}}(\hbar \omega )^{-2}(\varepsilon -{\frac{1}{2}}\hbar \omega
) $ \\ 
${N_{0}/N}$ & $1-({\frac{T}{T_{c}}})^{3}$ & $1-({\frac{T}{T_{c}}})^{5/2}$ & $%
1-({\frac{T}{T_{c}}})^{2}$ \\ 
$T_{c}$ & ${\frac{\hbar \omega }{k_{B}}}[{\frac{N}{\zeta (3)}}]^{1/3}$ & ${%
\frac{\hbar \omega }{k_{B}}}\left[ {\frac{(2\pi )^{1/2}N}{\zeta (3/2)}}{%
\frac{x_{0}}{L}}\right] ^{2/5}$ & ${\frac{\hbar \omega }{k_{B}}}\left[ {%
\frac{2\pi N}{\zeta (2)}}{\frac{x_{0}^{2}}{L^{2}}}\right] ^{1/2}$ \\ 
${U/Nk_{B}T}$ & ${\frac{3}{2}}{\frac{\hbar \omega }{k_{B}T}}+3({\frac{T}{%
T_{c}}})^{3}\frac{g_{4}(z_{1})}{\zeta (3)}$ & ${\frac{\hbar \omega }{k_{B}T}}%
+{\frac{5}{2}}({\frac{T}{T_{c}}})^{5/2}{\frac{g_{7/2}(z_{1})}{\zeta (5/2)}}$
& ${\frac{1}{2}}{\frac{\hbar \omega }{k_{B}T}}+2({\frac{T}{T_{c}}})^{2}{%
\frac{g_{3}(z_{1})}{\zeta (2)}}$ \\ 
$C\hbox{v}/Nk_{B}$ $(T<T_{c})$ & $12\left( \frac{T}{T_{c}}\right) ^{3}\frac{%
\zeta (4)}{\zeta (3)}$ & $\frac{35}{4}\left( \frac{T}{T_{c}}\right) ^{5/2}%
\frac{\zeta (7/2)}{\zeta (3/2)}$ & $6\left( \frac{T}{T_{c}}\right) ^{2}\frac{%
\zeta (3))}{\zeta (2)}$ \\ 
$C\hbox{v}/Nk_{B}$\ $(T>T_{c})$ & $12\frac{g_{4}(z_{1})}{g_{3}(z_{1})}-9%
\frac{g_{3}(z_{1})}{g_{2}(z_{1})}$ & $\frac{35}{4}\frac{g_{7/2}(z_{1})}{%
g_{5/2}(z_{1})}-\frac{25}{4}\frac{g_{5/2}(z_{1})}{g_{3/2}(z_{1})}$ & $6\frac{%
g_{3}(z_{1})}{g_{2}(z_{1})}-4\frac{g_{2}(z_{1})}{g_{1}(z_{1})}$ \\ 
$\Delta C_{\hbox{v}}/Nk_{B}$ & $9{\frac{\zeta (3)}{\zeta (2)}}\simeq 6.57$ & 
${\frac{25}{4}}{\frac{\zeta (5/2)}{\zeta (3/2)}}\simeq 3.20$ & $0$ \\ 
$PV$ & ${\frac{1}{3}}(U-{\frac{3}{2}}N\hbar \omega )$ & ${\frac{2}{5}}%
(U-N\hbar \omega )$ & ${\frac{1}{2}}(U-{\frac{1}{2}}N\hbar \omega )$ \\ 
$S/Nk_{B}$ & $4({\frac{T}{T_{c}}})^{3}{\frac{g_{4}(z_{1})}{\zeta (3)}}-%
\hbox{ln}z_{1}$ & ${\frac{7}{2}}({\frac{T}{T_{c}}})^{5/2}{\frac{%
g_{7/2}(z_{1})}{\zeta (5/2)}}-\hbox{ln}z_{1}$ & $3({\frac{T}{T_{c}}})^{2}{%
\frac{g_{3}(z_{1})}{\zeta (2)}}-\hbox{ln}z_{1}$ \\ \hline
\end{tabular}
\end{center}

\begin{quotation}
{\bf Table 1.} Thermodynamic quantities, as defined in text, for a
3D boson gas trapped by $\delta =3,\,2,1$ harmonic oscillators, with the
oscillator length parameter $x_{0}\equiv \left( {\hbar /m\omega }\right)
^{1/2}$.
\end{quotation}

An ideal Bose gas in $d$-dimensional space trapped by $\delta \leq d$
harmonic oscillators has its geometric dimensionality effectively reduced.
The BEC temperature expression (\ref{Tc}) for a trapped noninteracting Bose
gas shows that BEC can occur if and only if$\ \left( d+\delta \right) /2>1$
as otherwise the term $g_{(d+\delta )/2}(1)$ diverges forcing $T_{c}$ to
vanish. Thus BEC is possible in 2D provided $\delta \geq 1$.

Experiments with dilute boson gases confined in the realistic confining
potentials of opto-magnetic traps in the region of small oscillations where
(BEC) has been observed, can be viewed as a Bose gas in 3D with $\delta =3.$
Table 2 illustrates some parameters in bosonic vapor systems where BEC has
thus far been observed.

\begin{center}
\begin{tabular}{|c|c|c|c|c|c|c|c|}
\hline
{\small Boson} & $_{37}^{87}${\small Rb} & $_{11}^{27}${\small Na} & $%
_{3}^{7}${\small Li} & $_{1}^{1}${\small H} & $_{37}^{85}${\small Rb} & $%
_{2}^{4}${\small He} & $_{19}^{41}${\small K} \\ 
{\small Year/Ref.} & {\small 1995 \cite{Ander}} & {\small 1995 \cite{Davis}}
& {\small 1995 \cite{Bradley}} & {\small 1998 \cite{Fried}} & {\small 2000 
\cite{Cornish}} & {\small 2001 \cite{Pereira}} & {\small 2001 \cite{Mondugno}%
} \\ \hline
$N$ & $2\times 10^{4}$ & $5\times 10^{5}$ & $2\times 10^{5}$ & {\small -} & -
& $8\times 10^{6}$ & - \\ \hline
$N_{0}$ & $2\times 10^{3}$ & {\small -} & {\small -} & $10^{9}$ & $10^{4}$ & 
$5\times 10^{5}$ & $10^{4}$ \\ \hline
$T_{c}${\small \ }$(\mu K)$ & $0.17$ & $2$ & $0.4$ & $50$ & $0.015$ & $4.7$
& $0.16$ \\ \hline
$n${\small \ }$(cm^{-3})$ & $2.5\times 10^{12}$ & $1.5\times 10^{14}$ & $%
2\times 10^{12}$ & $4.8\times 10^{15}$ & $1\times 10^{12}$ & $3.8\times
10^{13}$ & $6\times 10^{11}$ \\ \hline
\end{tabular}
\end{center}

\begin{quotation}
{\bf Table 2.} Some experimental parameters associated with trapped bosonic
gases in which BEC has been observed to date, $N$ and $N_{0}$ being the
number of atoms in the initial cloud and in the condensate, respectively; $%
T_{c}$ the BEC transition temperature; $n$ the boson number density.
\end{quotation}

\section{Trapped fermions}

Finally, consider a system of $N$ noninteracting fermions in $d$ dimensions
trapped by $\delta $ $(\leq d)$ mutually perpendicular harmonic oscillators,
and otherwise free in the remaining $d-\delta $ directions. Since $\left[
e^{\beta \{\varepsilon -\mu (T)\}}+1\right] ^{-1}%
\smash {\
\mathop{\relbar\joinrel\longrightarrow}\limits_{T\to 0}\ \ }\theta \left(
E_{F}-\varepsilon \right) $, with\ $\mu \left( 0\right) \equiv $\ $%
E_{F}\equiv \hbar ^{2}k_{F}^{2}/2m$ the Fermi energy, $k_{F}$ being the
Fermi wavenumber, we see\ from (\ref{Num}) with $a=+1$ that 
\begin{eqnarray}
&&N\smash {\ \mathop{\relbar\joinrel\longrightarrow}\limits_{T\to 0}\ \ }%
\left[ 2A_{d+\delta }{\LARGE /}\left( {d+\delta }\right) \Gamma \left[
(d+\delta )/2\right] \right] \left( E_{F}-\hbar \omega \delta /2\right)
^{(d+\delta )/2}  \label{N2} \\
&\simeq &\left[ 2A_{d+\delta }{\LARGE /}\left( {d+\delta }\right) \Gamma %
\left[ (d+\delta )/2\right] \right] E_{F}^{(d+\delta )/2}  \nonumber
\end{eqnarray}
where in the last step we neglected $\hbar \omega \delta /2$ compared with $%
E_{F}$. The fermion number density with ${\tt s}=1/2$, if \ $\delta =0$ is
obtained \cite{Ariel, casas1}, substituting (\ref{Ad}) in (\ref{N2}), as 
\begin{equation}
n\equiv \frac{N}{L^{d}}=\frac{k_{F}^{d}}{2^{d-2}\pi ^{d/2}d\;\Gamma (d/2)},
\label{eq:ndd}
\end{equation}
which reduces to the familiar results $n=2k_{F}/\pi ,$ $k_{F}^{2}/2\pi $ and 
$k_{F}^{3}/3\pi ^{2}$ for $d=1,$ $2$ and $3$, respectively.

Recalling that $-{Li_{\sigma }(-z)}\equiv f_{\sigma }(z)$ which are the FD
integrals, the internal energy from (\ref{u-3}) can be expressed as 
\begin{equation}
\frac{U(T,V)}{Nk_{B}T}=\left[ \beta {\frac{\hbar \omega \delta }{2}}+{\frac{%
d+\delta }{2}}\frac{{f_{\left( d+\delta \right) {/2}+1}(z_{1})}}{{f_{\left(
d+\delta \right) {/2}}(z_{1})}}\right] .  \label{uf}
\end{equation}
Using (\ref{N2}) and the asymptotic expansion for $f_{d/2}(z)$ for $%
T\rightarrow 0$ (Ref. \cite{Ariel}, App.\ B), (\ref{uf}) becomes 
\begin{equation}
\frac{U(T)-N\hbar \omega \delta /2}{Nk_{B}T_{F}}\smash {\
\mathop{\relbar\joinrel\longrightarrow}\limits_{T\to 0}\ \ }\frac{({d+\delta
)}}{({d+\delta +2)}}\left[ {1\ +\ }({d+\delta +2)}\frac{{\pi }^{2}}{12}%
\left( \frac{T}{T_{F}}\right) ^{2}\right] .  \label{u0}
\end{equation}
From ($\ref{CV}$) the specific heat as $T\rightarrow 0$ is then 
\begin{equation}
\frac{C_{V}(T)}{Nk_{B}}\smash {\
\mathop{\relbar\joinrel\longrightarrow}\limits_{T\to 0}\ \ }({d+\delta )}%
\frac{{\pi }^{2}}{6}\left( \frac{T}{T_{F}}\right) ,  \label{cv0}
\end{equation}
while the entropy $S=\int_{0}^{T}dT^{^{\prime }}C_{V}(T^{^{\prime
}})/T^{^{\prime }}$ becomes 
\begin{equation}
S/Nk_{B}\smash {\ \mathop{\relbar\joinrel\longrightarrow}\limits_{T\to 0}\ \
}({d+\delta )}\frac{{\pi }^{2}}{6}\left( \frac{T}{T_{F}}\right)
\end{equation}
which again is in agreement with the third law of thermodynamics. These
results are reflected in Fig. 3. Table 3 summarizes results for 3D fermions
with $\delta =1,\,2,\,3$ harmonic oscillators.

\begin{center}
\begin{tabular}{|c||c|c|c|}
\hline
$\delta $ & $3$ & $2$ & $1$ \\ \hline\hline
$N$ & ${\frac{2}{3}}(\hbar \omega )^{-3}E_{F}^{3}$ & ${\frac{4}{5}}({\frac{%
mL^{2}}{{2\pi \hbar ^{2}}}})^{1/2}(\hbar \omega )^{-2}E_{F}^{5/2}$ & $({%
\frac{mL^{2}}{{2\pi \hbar ^{2}}}})(\hbar \omega )^{-1}E_{F}^{2}$ \\ 
$U/Nk_{B}T$ & ${\frac{3\hbar \omega }{2k_{B}T}}+\frac{3f_{4}(z_{1})}{%
f_{3}(z_{1})}$ & $\frac{\hbar \omega }{k_{B}T}+\frac{5f_{7/2}(z_{1})}{%
2f_{3/2}(z_{1})}$ & $\frac{\hbar \omega }{2k_{B}T}+\frac{2f_{3}(z_{1})}{%
f_{2}(z_{1})}$ \\ 
${C_{V}/Nk_{B}}$ & $12\frac{f_{4}(z_{1})}{f_{3}(z_{1})}-9\frac{f_{3}(z_{1})}{%
f_{2}(z_{1})}$ & $\frac{35}{4}\frac{f_{7/2}(z_{1})}{f_{5/2}(z_{1})}-\frac{25%
}{4}\frac{f_{5/2}(z_{1})}{f_{3/2}(z_{1})}$ & $6\frac{f_{3}(z_{1})}{%
f_{2}(z_{1})}-4\frac{f_{2}(z_{1})}{f_{1}(z_{1})}$ \\ 
${PV/Nk_{B}T}$ & $\frac{f_{4}(z_{1})}{f_{3}(z_{1})}$ & $\frac{f_{7/2}(z_{1})%
}{f_{5/2}(z_{1})}$ & $\frac{f_{3}(z_{1})}{f_{2}(z_{1})}$ \\ 
$S/Nk_{B}$ & $\frac{4f_{4}(z_{1})}{f_{3}(z_{1})}-\hbox{ln}z_{1}$ & $\frac{%
7f_{7/2}(z_{1})}{2f_{5/2}(z_{1})}-\hbox{ln}z_{1}$ & $\frac{f_{3}(z_{1})}{%
f_{2}(z_{1})}-\hbox{ln}z_{1}$ \\ \hline
\end{tabular}
\end{center}

\begin{quotation}
{\bf Table 3.} Thermodynamic quantities, as defined in text, for a
3D fermion gas trapped by $\delta =1,\,2,\,3$ harmonic oscillators.
\end{quotation}

\section{Conclusions}

After constructing the grand potential, thermodynamic properties were
determined along with the densities of states of ideal boson and fermion
gases in $d$ dimensions trapped by $\delta $ mutually perpendicular harmonic
(HO) oscillators. Trapping maps the system into a free gas with a new
dimensionality increased by the number of trapping oscillators,
specifically, $d\rightarrow d+\delta $, and renormalizes the particle masses 
$m\rightarrow m^{\ast }$ according to (\ref{mequiv}). In particular, we
detailed how 3D boson and fermion gases trapped by 1, 2 or 3
mutually-perpendicular HO\ wells {\it map} into a free gas in 4, 5 and 6
dimensions, respectively. Also, we found that in a trapped boson gas
Bose-Einstein condensation with critical temperature $T_{c}\neq 0$ occurs if
and only if $d+\delta >2$ so that for $\delta \geq 1$, $d$ need not be
restricted to $d>2$.

\section*{Acknowledgments}

We acknowledge partial support from UNAM-DGAPA-PAPIIT (M\'{e}xico) \#
IN102198 and CONACyT (M\'{e}xico) \# 27828 E. M.G. acknowledge fellowship from CONACyT (M\'{e}xico).

\bigskip


\begin{figure}[tbh]
\centerline{\psfig{file=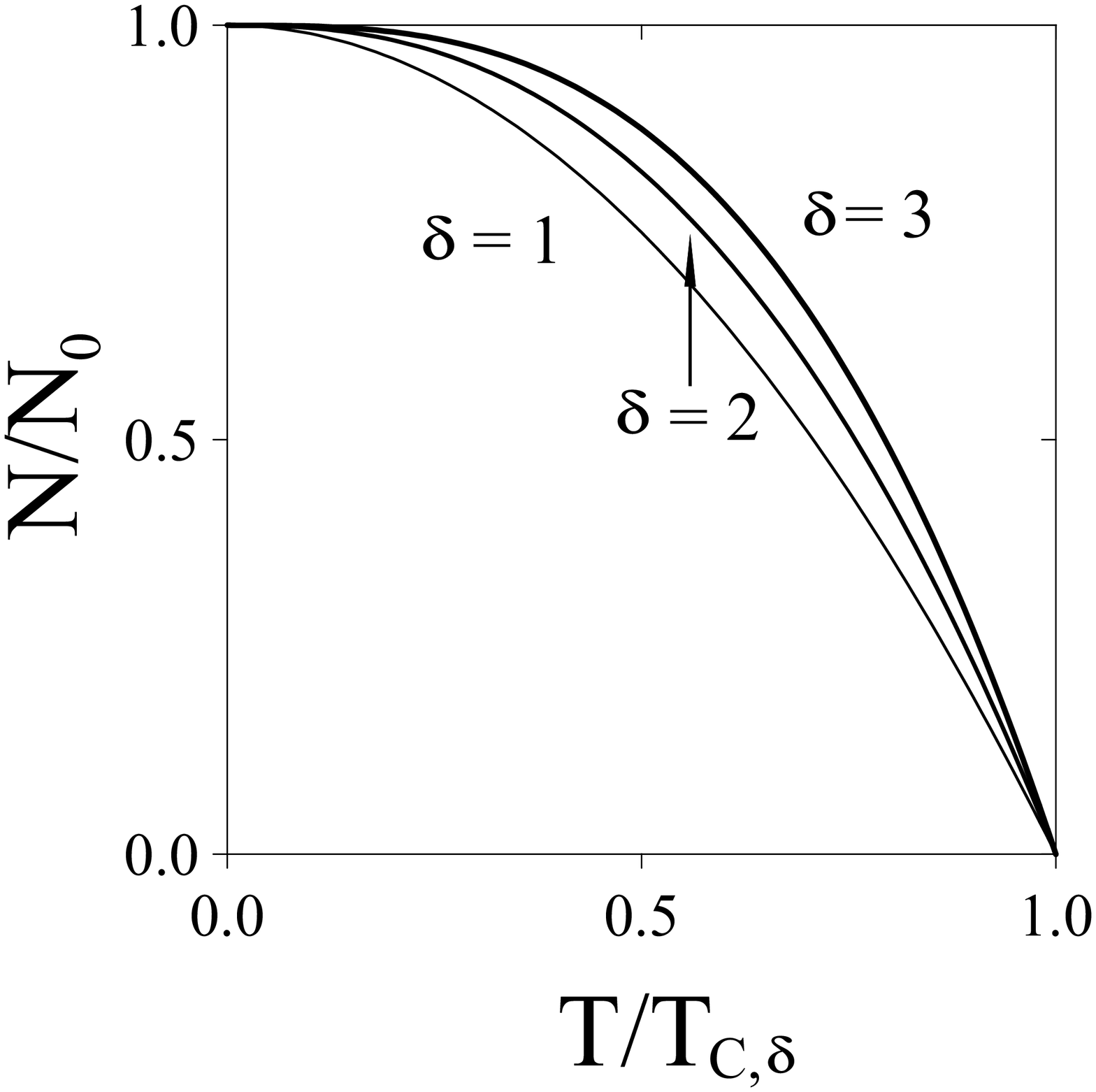,height=3.0in,width=3.0in}}
\caption{Condensate fraction for a 3D boson gas trapped by $\delta =1$, 2 or 3 harmonic oscillators.
}
\end{figure}

\begin{center}
\begin{figure}[tbh]
\begin{minipage}[b]{3.0in}
\centerline{\psfig{file=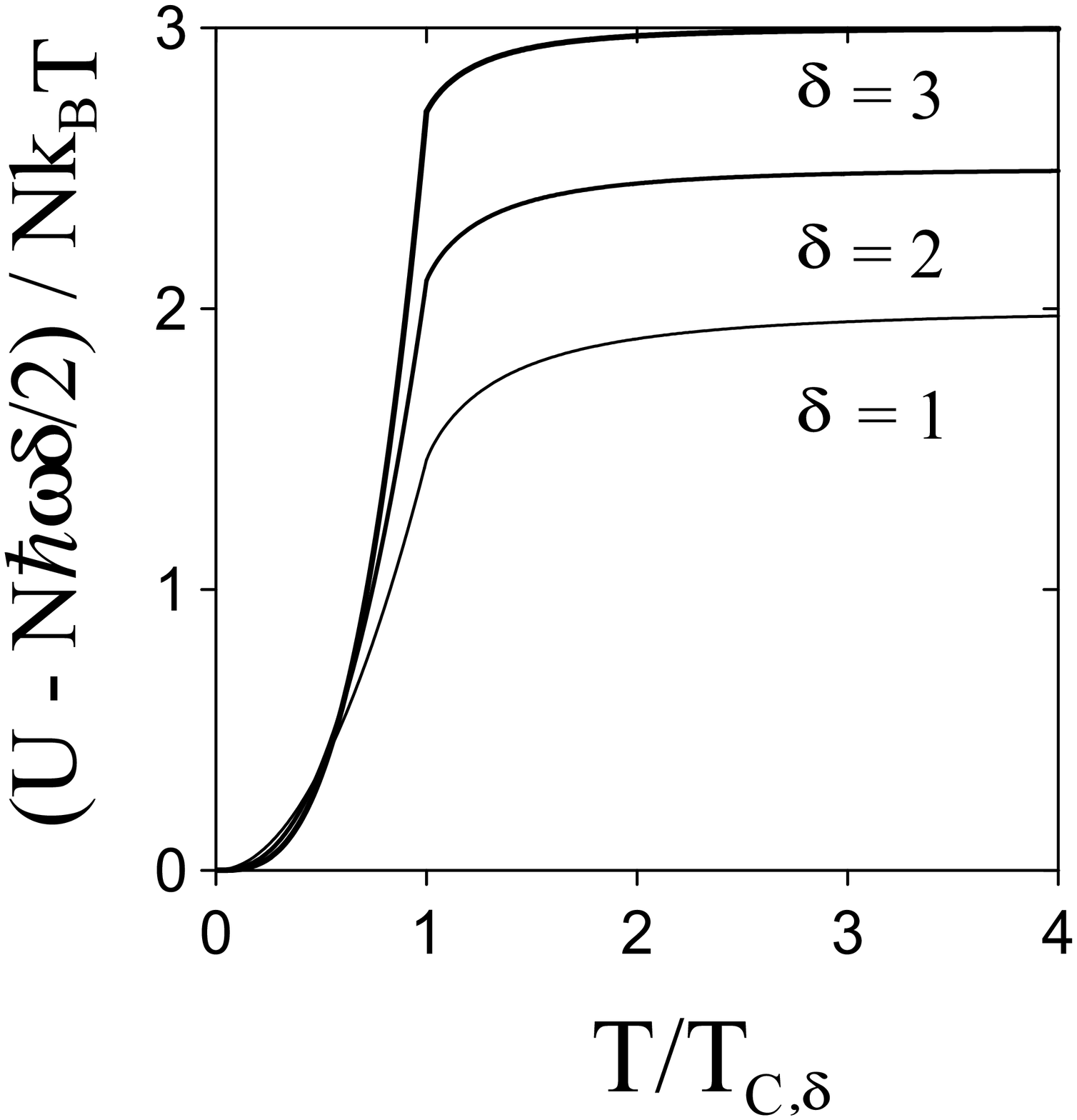,height=3.0in,width=3.0in}}
\end{minipage}
\vspace{1.0cm}
\begin{minipage}[b]{3.0in} 
\centerline{\psfig{file=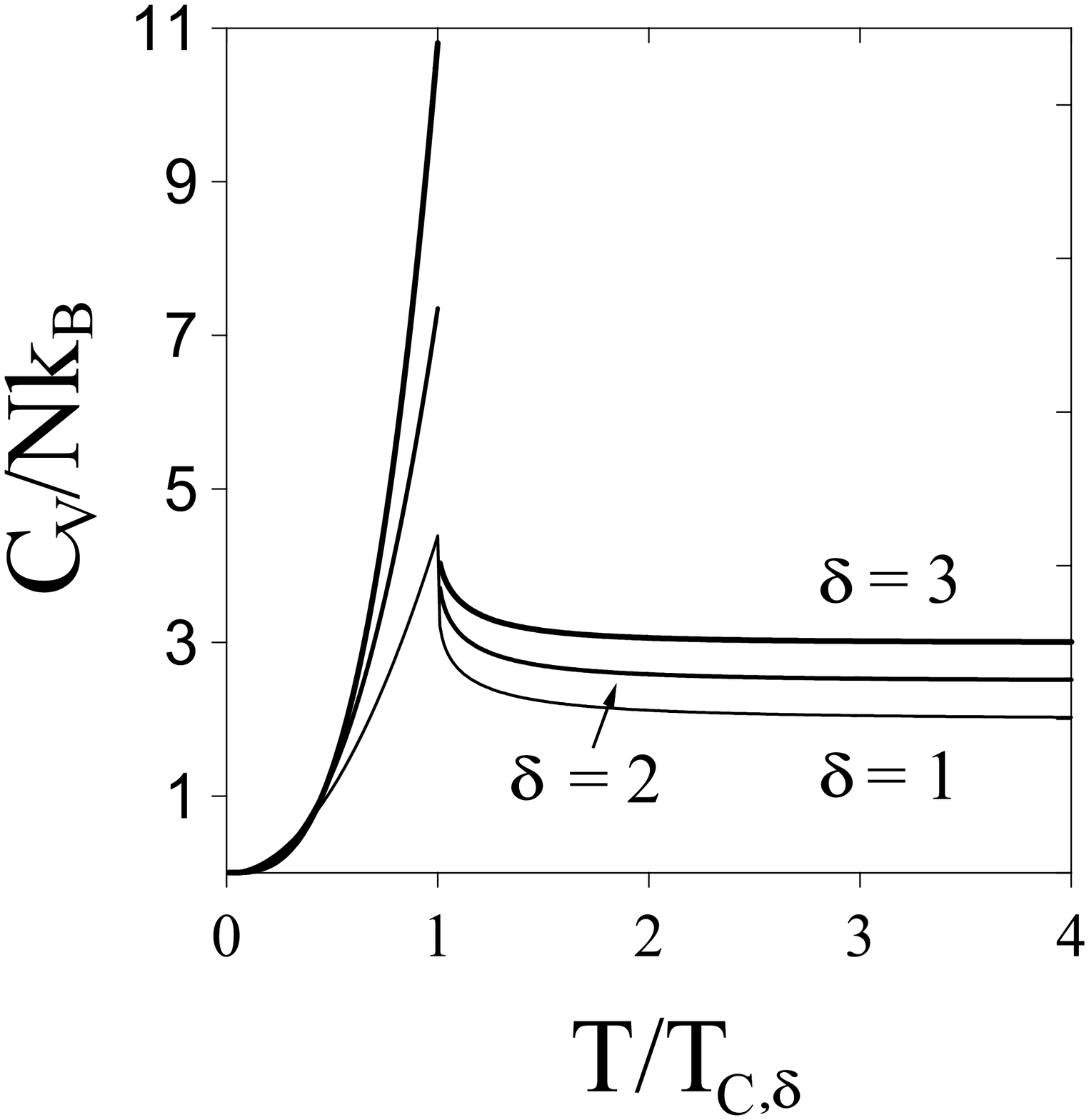,height=3.0in,width=3.0in}}
\end{minipage}
\vspace{1.0cm} 
\begin{minipage}[b]{3.0in}
\centerline{\psfig{file=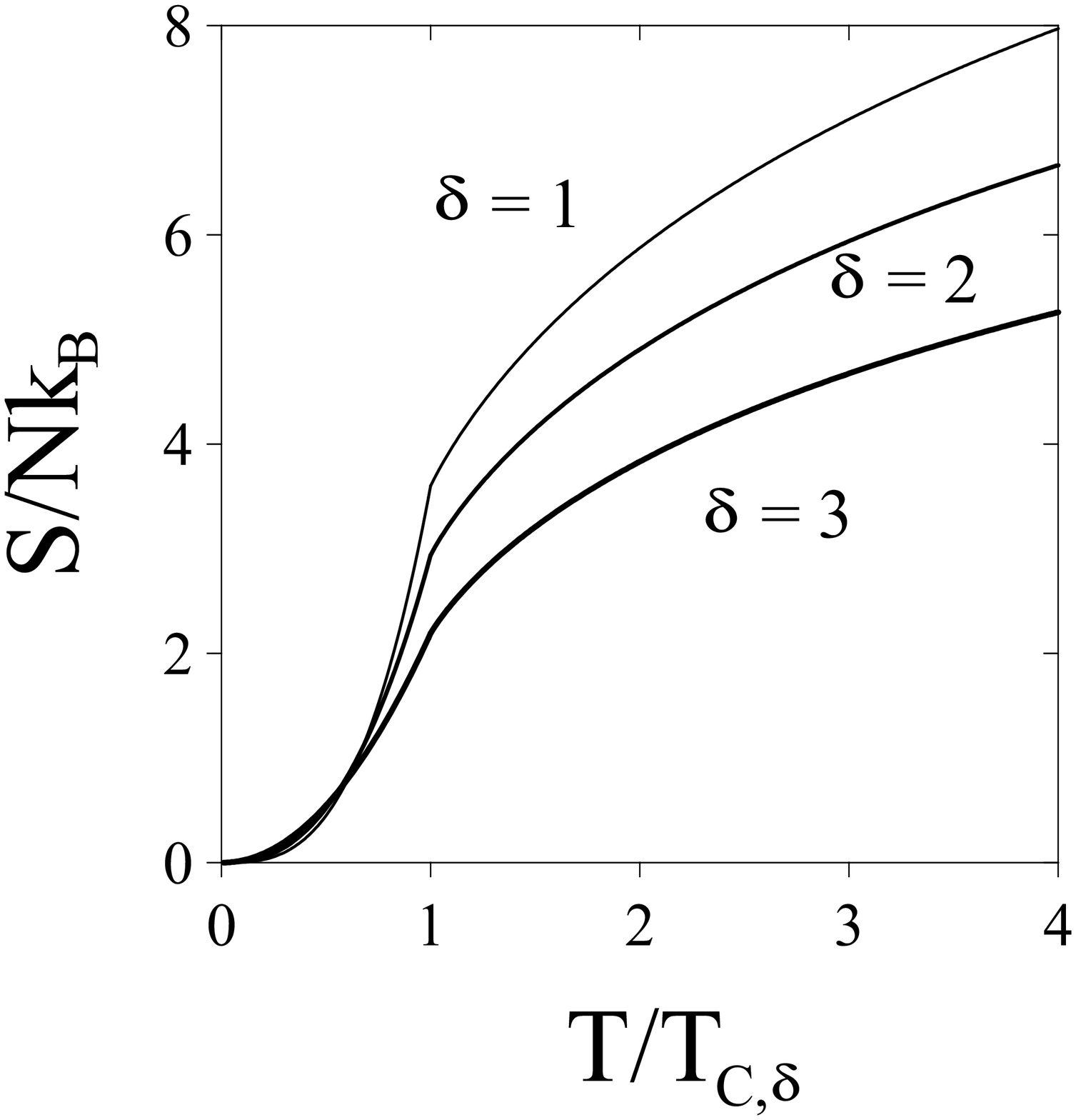,height=3.0in,width=3.0in}}
\end{minipage}\hfill
\begin{minipage}[b]{3.0in} 
\centerline{\psfig{file=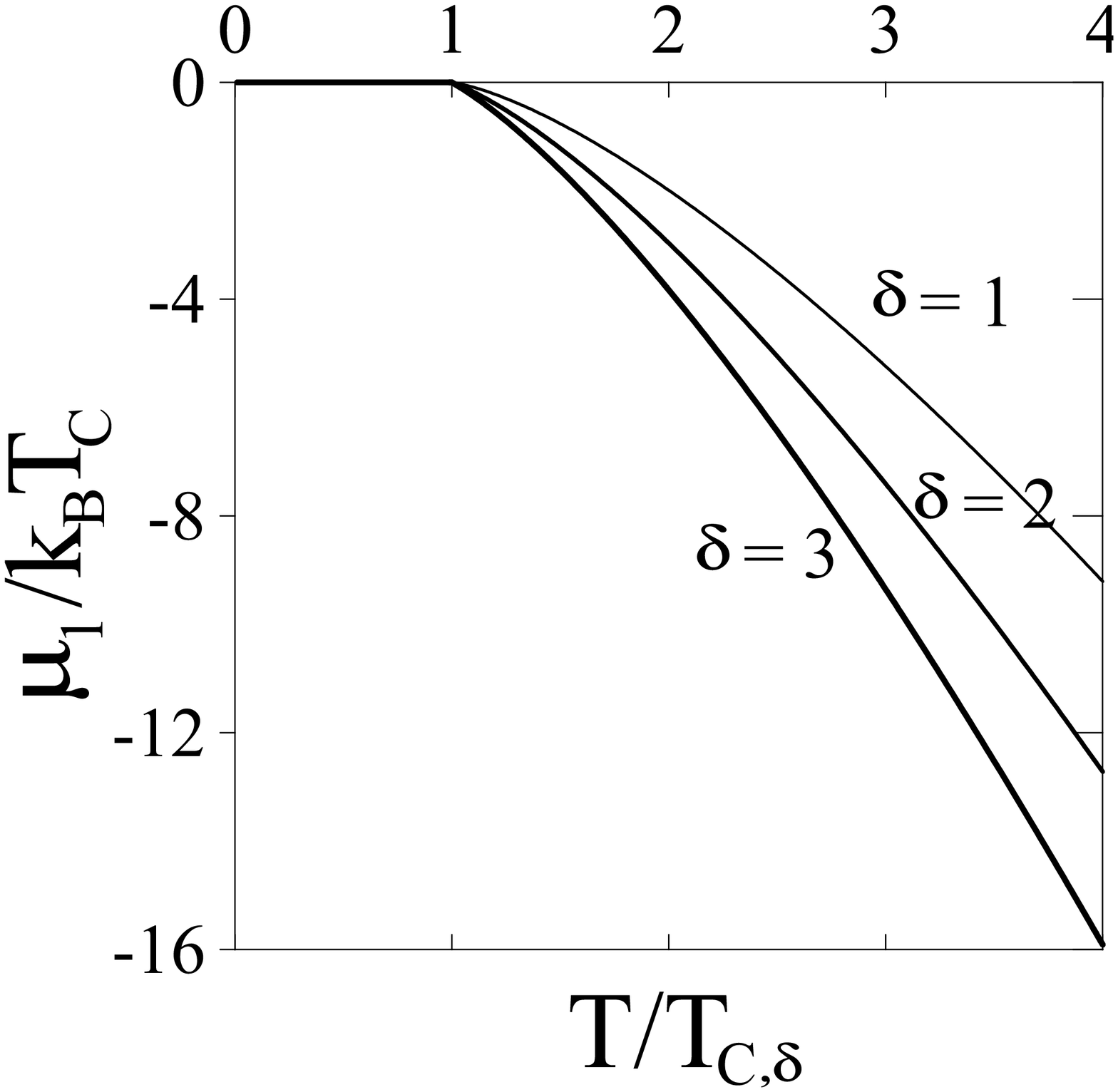,height=3.0in,width=3.0in}}
\end{minipage}
\label{fig:tau2} 
\vspace{-1.0cm}
\caption{Thermodynamic variables as functions of temperature $T$, as defined in text,\ for a 3D boson gas trapped by $\delta =1$, 2 or 3 harmonic oscillators.
}
\end{figure}
\end{center}

\begin{figure}[tbh]
\begin{minipage}[b]{3.0in}
\centerline{\psfig{file=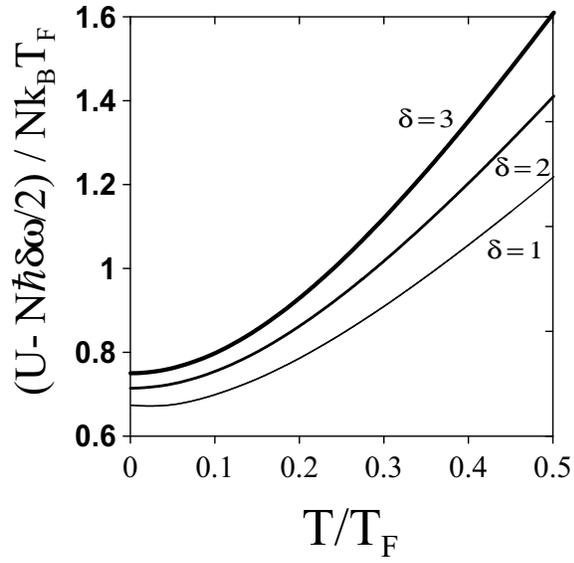,height=3.0in,width=3.0in}}
\end{minipage}
\vspace{1.0cm} 
\begin{minipage}[b]{3.0in} 
\centerline{\psfig{file=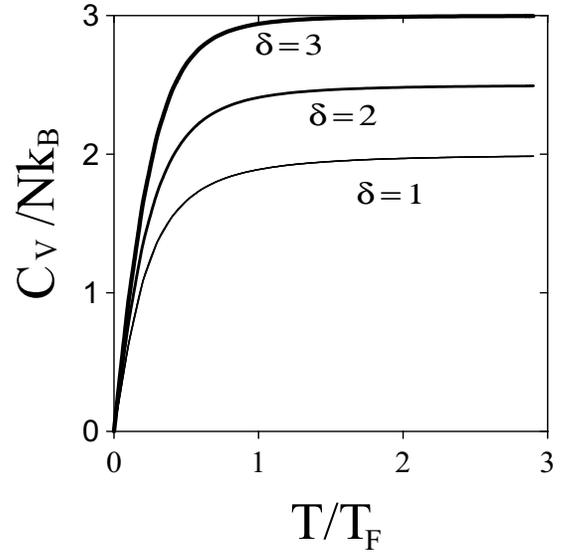,height=3.0in,width=3.0in}}
\end{minipage}
\vspace{1.0cm} 
\begin{minipage}[b]{3.0in}
\centerline{\psfig{file=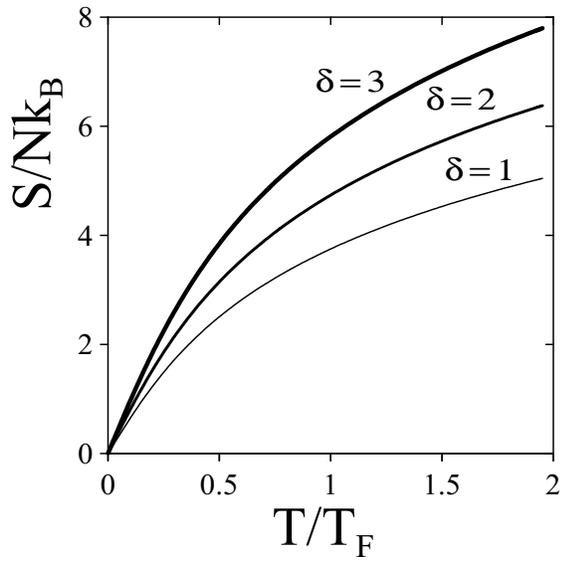,height=3.0in,width=3.0in}}
\end{minipage}\hfill
\begin{minipage}[b]{3.0in} 
\centerline{\psfig{file=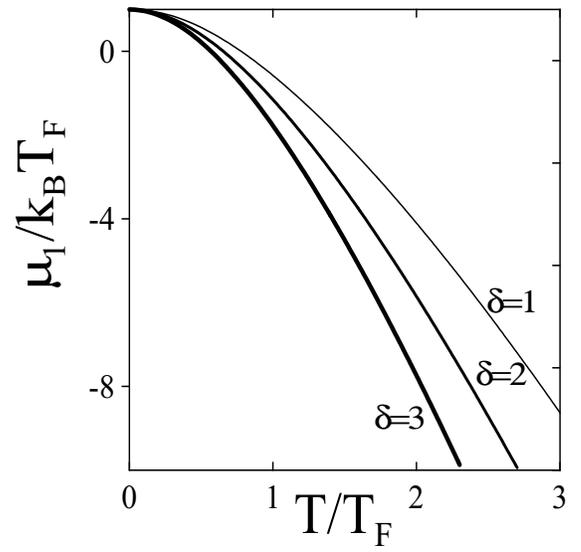,height=3.0in,width=3.0in}}
\end{minipage}
\label{fig:tau2} 
\vspace{-1.0cm}
\caption{Same as Fig. 2 but for a fermion gas.}
\end{figure}

\end{document}